\documentclass[a4paper,11pt]{article}
\pdfoutput=1 

\usepackage{jinstpub} 

\usepackage{cleveref}
\crefname{figure}{figure}{figures}
\usepackage{caption}
\usepackage{multirow}
\usepackage{subcaption}
\usepackage{alltt}
\usepackage{epstopdf}
\usepackage{setspace}
\usepackage{lineno}
\usepackage{dcolumn}
\usepackage[perpage]{footmisc}
\usepackage{amsmath,amsfonts,amssymb,wasysym,ifsym}
\usepackage{color}
\usepackage{MnSymbol}
\usepackage{verbatim}
\usepackage{upgreek}
\usepackage{hyperref}
\usepackage{bm}
\usepackage{textcomp}
\usepackage{graphicx}
\usepackage[T1]{fontenc}
\usepackage{libertine}
\usepackage{xspace}

\newcommand{\myvec}[1]{\begin{tabular}{c} #1 \end{tabular}}
\usepackage{footnote}
\makesavenoteenv{tabular}
\makesavenoteenv{table}

\usepackage{lineno}
\title{Production and validation of scintillating structural components from low-background Poly(ethylene naphthalate)}
\makeatletter
\newcommand\footnoteref[1]{\protected@xdef\@thefnmark{\ref{#1}}\@footnotemark , }
\makeatother

\author[c]{Y.~Efremenko,}
\author[b]{M.~Febbraro,}
\author[a]{F.~Fischer,}
\author[a]{M.~Guitart Corominas,}
\author[h]{K.~Gusev,}
\author[c]{B.~Hackett,}
\author[d]{C.~Hayward,}
\author[e]{R.~Hod\'ak,}
\author[h]{P.~Krause,}
\author[a]{B.~Majorovits,}
\author[a]{L.~Manzanillas,} 
\author[d]{D.~Muenstermann,}

\author[f]{M.~Pohl,}
\author[f]{R.~Rouhana,}
\author[b]{D.~Radford,} 
\author[e]{E.~Rukhadze,}
\author[h]{N.~Rumyantseva,}
\author[f]{I.~Schilling,}
\author[h]{S.~Schoenert,} 
\author[a]{O.~Schulz,}
\author[h]{M.~Schwarz,}
\author[e]{I.~Štekl,}
\author[g]{M.~Stommel,}
\author[f]{J.~Weingarten,}

\author[i]{E.~Hoppe,}
\author[i]{I.~Arnquist,}
\author[i]{K.~Hobbs,}
\author[i]{A.~French,}
\author[i]{M.~diVacri,}

\author[j]{M.~Laubenstein,}

\author[k]{G.~Zuzel}

\affiliation[a]{Max-Planck-Institut  f\"ur  Physik,  80805  Munich,  Germany}
\affiliation[b]{Oak Ridge National Laboratory, Oak Ridge, Tennessee 37830}
\affiliation[c]{Department of Physics and Astronomy, University of Tennessee, Knoxville, Tennessee 37916}
\affiliation[d]{Department of Physics, Lancaster University, Lancaster}
\affiliation[e]{Institute of Experimental and Applied Physics, Czech Technical University in Prague, CZ-11000 Prague, Czech Republic}
\affiliation[f]{Technische Universität Dortmund, Dortmund}
\affiliation[g]{Leibniz-Institut für Polymerforschung Dresden e.V., 01069 Dresden, Germany}
\affiliation[h]{Physik Department, Technische Universität, München}
\affiliation[i]{Pacific Northwest National Laboratory, Richland, WA, USA}
\affiliation[j]{INFN, Laboratori Nazionali del Gran Sasso, 67100 Assergi, Italy}
\affiliation[k]{Institute of Physics, Jagiellonian University, 31-007 Cracow, Poland}

\emailAdd{ffischer@mpp.mpg.de, manzanil@mpp.mpg.de}
\abstract{Poly Ethylene Naphthalate (PEN) is an industrial polymer plastic
which 
is investigated as a low background, 
transparent, scintillating and wavelength shifting structural material.
PEN scintillates in the blue region and has excellent mechanical properties both at room and cryogenic temperatures.  Thus, it is an ideal candidate for active structural components in 
experiments for the search of rare events like neutrinoless double-beta decay or dark matter recoils. 
Such optically active structures 
improve the identification and rejection 
efficiency of 
backgrounds events, 
like this improving the sensitivity of experiments. 
This paper reports on the production of radiopure and 
transparent PEN plates 

These structures can be used to mount germanium detectors operating in cryogenic liquids (LAr, LN). Thus, as first application PEN holders will be used to mount the Ge detectors in the \textsc{Legend}-$200$ experiment. The whole process from cleaning the raw material to testing the PEN active components under final operational conditions is reported. { \color{red} \it \footnotesize }}

\begin{document}
\maketitle

\section{Introduction}
\label{sec:intro}


The next generation of low-background experiments is presently being prepared. 
These ton-scale experiments demand quasi background-free environments in their region of interest to achieve the scientific goals. For this,
new innovations in addition to the use of ultra-pure materials are required. Active shields have been used in experimental setups to actively identify and veto background events. This approach has proven to be very efficient as demonstrated by the \textsc{Gerda} experiment in the search for neutrinoless double beta decay ($0\nu\beta\beta$) \cite{GERDA:2020xhi}.
In \textsc{Gerda}, one of the predecessors of \textsc{Legend}-$200$ \cite{LEGEND:2021bnm}, the Germanium (Ge) detectors were operated in liquid argon\,(LAr) \cite{Agostini:2017hit}. The LAr not only serves to cool down the detectors and as ultra-pure passive shield, but also acts as an active veto system against external radiation thanks to its scintillation properties. 
The scintillation photons from LAr have a wavelength of $\approx\,127$\,nm (vacuum ultraviolet region) \cite{Heindl:2010zz} which is not detectable by most standard photo sensors.
In the case of \textsc{Gerda}, a curtain of wavelength-shifting\,(WLS) optical fibers coated with tetraphenyl butadienne (TPB) 
was placed around the detector strings.
These fibers collect the photons from LAr scintillation, shift their wavelength and guide them towards an array of silicon photomultipliers (SiPMs) \cite{Janicsko-Csathy:2010uif}. This configuration allows the placement of photon detectors and their front-end electronics to be further away from the Ge detectors, decreasing the backgrounds induced by these components \cite{Agostini:2017hit}. 

The efficiency of detecting scintillation will suffer from absorption in optically inactive structural materials like detectors holders and other support structures. In \textsc{Gerda}, high purity silicon plates were used as holder material. These were partially between the Ge detectors. These prevent an efficient detection of LAr scintillation light from background events originating between the detectors.

Research is presently ongoing in developing high purity poly(ethylene $2,6$-naphthalate) or PEN $[$C$_{14}$H$_{10}$O$_{4}]_\text{n}$, as optically active structural material.
This research aims to replace optically inactive materials like copper, silicon and PTFE with PEN, improving the background detection efficiency.
PEN is known to scintillate in the deep blue region\,\cite{pennakamura}.
In addition, PEN also can shift the vacuum ultraviolet light to easily detectable blue light with a high efficiency \cite{Kuzniak:2018dcf,Boulay:2021njr}. 
Furthermore, it was recently shown that custom-made PEN structures have the necessary stability and resilience to be used as a holding structure for detectors \cite{Efremenko:2019xbs,Abt:2020pwk}.
In this context, the principles of \textsc{Gerda} have been adopted for \textsc{Legend}-$200$, including the LAr veto. In addition, active holding plates made from PEN will be used. This complements the active veto system by minimizing the absorption of photons on optically inactive materials and increasing the general veto efficiency due to the WLS and scintillation properties of PEN.

Section~\ref{sec:production} describes a methodology for producing active and low background components out of commercial available PEN granulate using compression injection molding technology. Section~\ref{sec:screening} reports the achieved radiopurity of this PEN production for usage in the \textsc{Legend}-$200$ experiment. In Section~\ref{sec:optical_properties}, preliminary results of the optical characterization are reported, while Section~\ref{sec:tests} describes the different tests performed in close to final operation conditions to validate the usage of PEN in the \textsc{Legend}-$200$ experiment. Finally, conclusions and future prospects are presented in Section~\ref{sec:conclusions}.   
\section{Low-background production of PEN structures}
\label{sec:production}

PEN raw material is commercially available in the form of white granules from Teijin-DuPont~\cite{dupontweb}. The material used for this \textsc{Legend}-$200$ production has the designation TN-$8065$ SC. The untreated PEN granulate had been previously screened for radio-impurities. Apart from a Bq/kg level contamination with $^{40}$K, low contamination levels at the $0.2$\,mBq/kg with $^{238}$U and $^{232}$Th were measured (see Table \ref{tab:PEN:rp}) \cite{Efremenko:2019xbs}. The high level of the $^{40}$K signal was attributed to surface contamination. 
To mitigate these, and other possible surface contamination, an extensive cleaning procedure was adopted. This procedure was applied as the first step of the production process.

After cleaning, the PEN granulate was used to produce discs by a specially optimized and controlled injection compression molding technique. Finally, the injection molded discs were used to machine the detector holders under a laminar flow hood. The detector holder geometry was adapted to the detector size, while minimizing the total weight. This section describes the details of the whole production chain.

\subsection{Cleaning procedure}
\label{sec:cleaning}

All parts that can come in contact with the granulate during the production process were cleaned with the following procedure: First, they were rinsed off in a flow box located in a cleanroom with $18$\,MOhm water (DI water) and then washed off with class\,$100$ cleanroom cloths dipped in VLSI\,grade $2$-propanol. The parts were then etched in a nitric acid solution ($2\,\%$ HNO$_3$ optima grade, $98\,\%$ DI water) for $24$\,h and rinsed off again with DI water of the same quality afterwards. Finally, the containers and sieves were cleaned separately for $15$\,minutes in an ultrasonic bath filled with DI water.

The PEN granulate is supplied by manufacturers in large plastic containers, containing $25$\,kg each. Cleaning of the PEN granulate was conducted in batches of $\approx 1$\,kg. Each batch underwent the following cleaning steps:
\begin{itemize}
    \item[1.] Fill PEN granulate into cleaned stainless sieve
    \item[2.] Rinse with $\approx 2$\,litres of DI water $18$\,MOhm inside flow box for two minutes
    \item[3.] The sieve with PEN is inserted into a PTFE container filled with $2.5$\,litres VLSI grade $2$-propanol
    \item[4.] PTFE container is placed inside ultrasonic bath containing $\approx 50$\,litres of DI water for 15 minutes
    \item[5.] Rinse PEN with DI water for about one minute
    \item[6.] Repeat steps $3$.-$6$. in separate ultrasound bath
    \item[7.] The sieve with PEN is inserted into PTFE container filled with $18$\,MOhm DI water
    \item[8.] PTFE container is placed inside ultrasonic bath containing $\approx 50$\,litres of DI water for 15 minutes
    \item[9.] Rinse PEN with DI water for about one minute
\end{itemize}

\begin{figure}[h]
    \centering
    \includegraphics[width=0.42\textwidth]{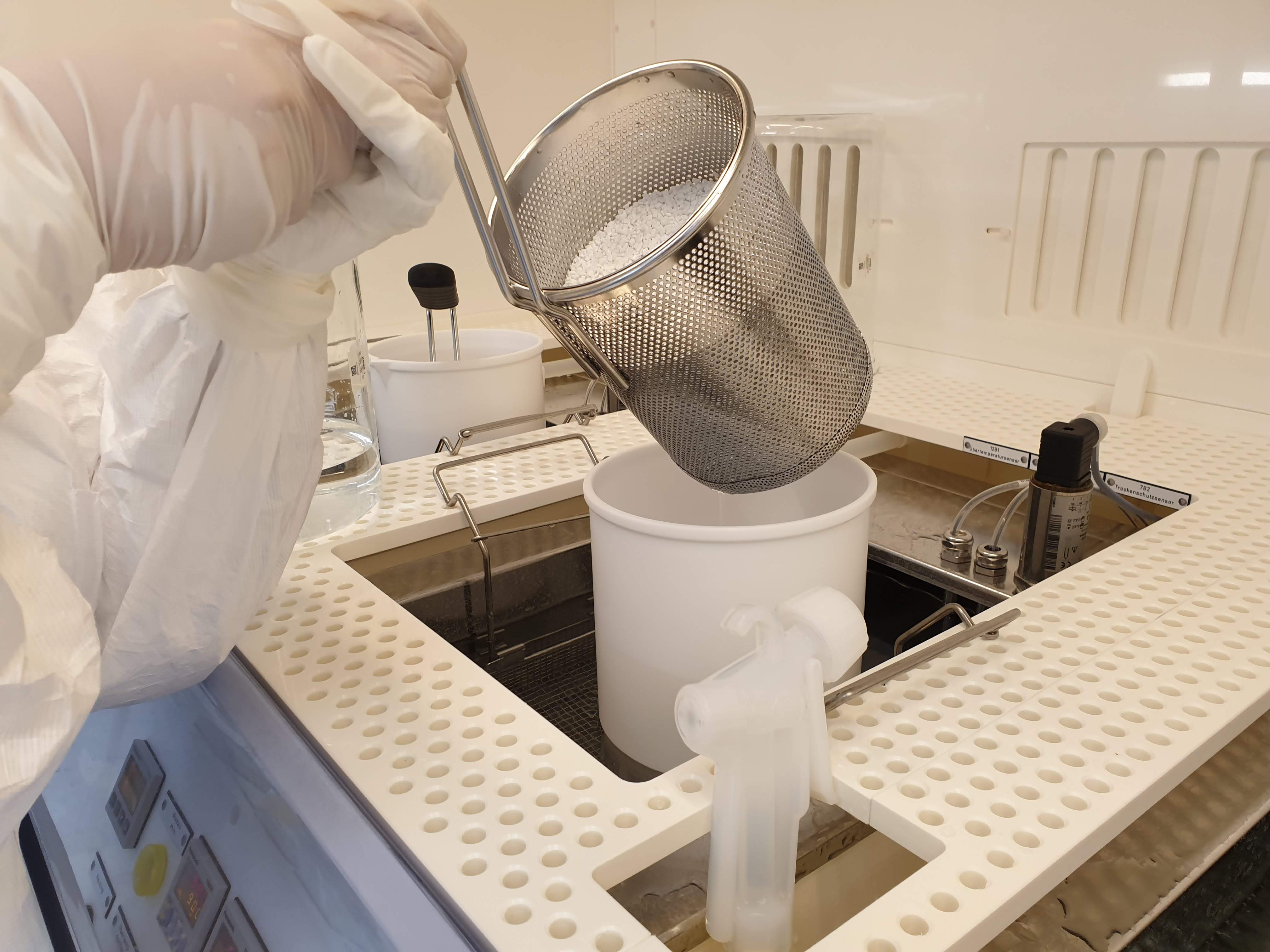}
    \includegraphics[width=0.56\textwidth]{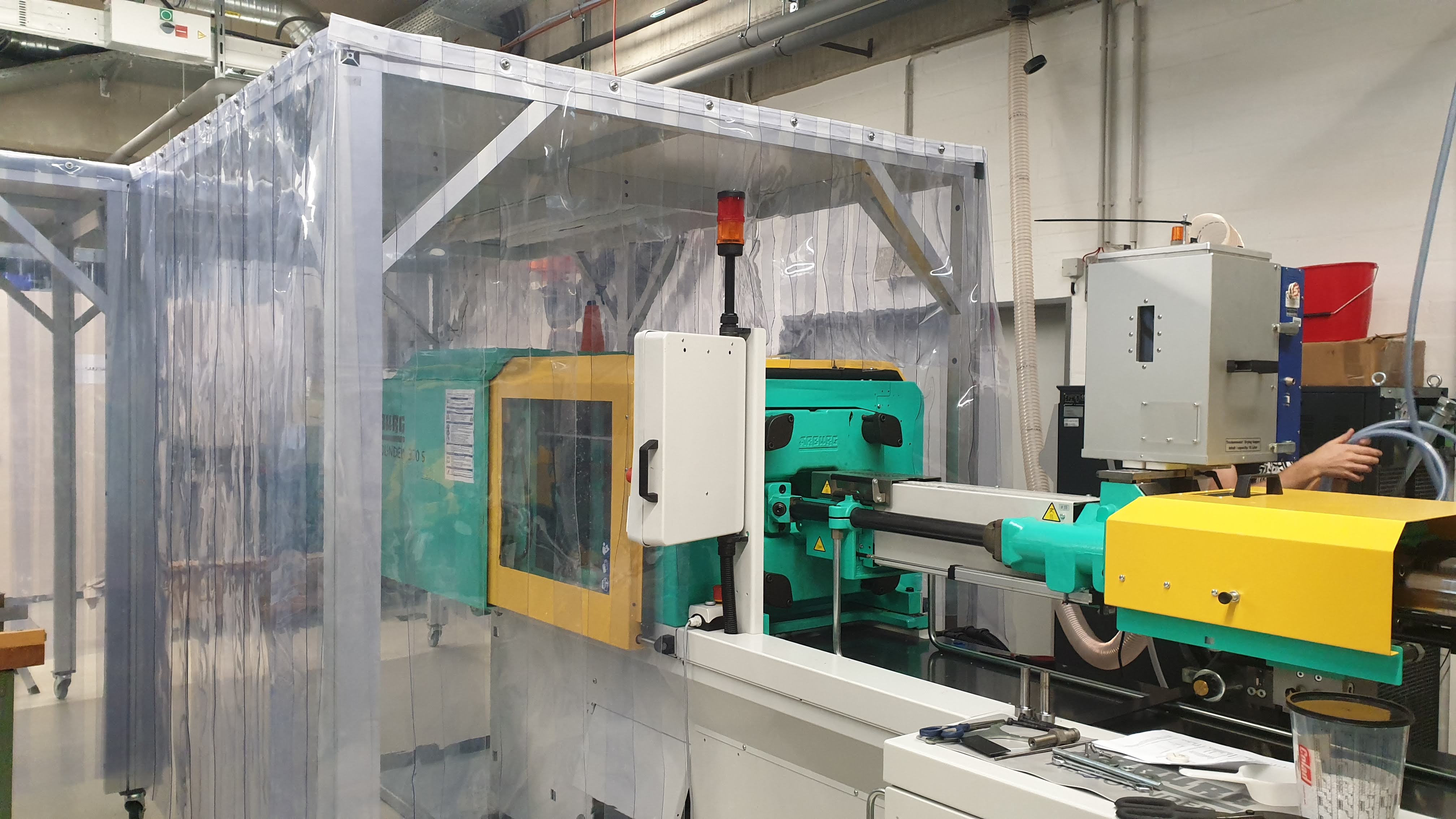}
    \caption{Left: PEN pellets in stainless-steel sieve during cleaning process. The PTFE container is placed in a basin filled with DI water for ultra sound cleaning. Right: Injection compression molding machine used for production of PEN discs. All open parts of the machine are located inside a clean room.}
    \label{fig:clean_machine}
\end{figure}

Finally, the granules were collected in a larger sieve in a separate basin. Most of the residual DI water is removed here.

The cleaned granulate was dried in two stages. In the first, $\approx 10$\,kg PEN granulate was filled into a sieve and baked out in a vacuum tank. An oil-free roughing pump was used. The tank was heated in order to accelerate the evaporation of the water residues using heating bands attached to the outside tank. The temperature was set to $60\,^\circ$C and after $8$\,hours, the pre-dried granulate was sealed in class\,$100$ plastic bags. These were then packed for transportation into another plastic bag of the same cleanliness.

\subsection{Production of scintillator-grade PEN discs}
\label{sec:production-process}

Injection compression molding was used to transform the cleaned PEN granulate to discs with a diameter of up to $35$\,cm and $1.5$\,mm thickness. The machine used here is an Arburg Allrounder $370$S (ARBURG GmbH, Lossburg, Germany) where the barrel and screw were replaced beforehand with new ones. The mold consisted of two new stainless steel plates (type $316$) with a mirror-like surface finishing (roughness $< 1$\,$\mu$m) which in industry is referred to as $1$P/$2$P or No.$8$ (ASTM A$480$ /A$480$M).

All containers and areas of the molding machine that came into direct contact with the granulate were cleaned using a nitric acid solution as described in Section~\ref{sec:cleaning}. The machine itself was located in a clean room (Figure~\ref{fig:clean_machine}, Right)  and was cleaned with $2$-propanol (VLSI grade) and 
Micro Cleaning Solution (Micro-$90^{\circledR}$) \cite{micro-90}. The machine's storage tank allowed to store the PEN pellets in nitrogen atmosphere. This container can be heated to remove the remaining moisture on the surface and inside the PEN pellets. In order to expel any radon from the granulate, nitrogen boiled off from a liquid nitrogen dewar was used to flush the storage volume while heating it for at least $6$\,hours to $80^\circ$C until the proper humidity was reached.

The optimal settings for the plasticizer unit of the molding machine were determined empirically. In first trials, the values recommended by the manufacturer were used and then optimized for the highest possible transparency of the outcome. For each produced PEN disc, approximately $125$\,cm$^3$ ($170$\,g) of material was used. The injection time was set to $2$\,s at a volume rate of $100$\,cm$^3$/s. The optimum temperature of the plasticizer unit of $290^\circ$C with a controlled temperature of the compression discs of $30^\circ$C was also established empirically. The compression takes place in two phases. At the beginning the steel plates of the mold are $5.5$\,mm apart and are closing at a speed of $3$\,mm/s until the distance between the plates reaches $2$\,mm. The speed is then reduced to $2$\,mm/s until the target compression pressure of $700$\,kN is reached, corresponding to a final disc thickness of $1.5$\,mm. After a cooling time of $40$\,s the discs could be removed manually without the need to use tools.
Some differences in the quality of the discs were notable from optical defects such as local crystallization and air inclusions. These defects occurred more frequently when the molding parameters were not fully optimized  at the beginning of a production series. 

After extracting the disc from the mold (see Figure~\ref{fig:holder_in_hand},~ Left), it was locked in a CNC (Computerized Numerical Control) milling machine in the clean room placed nearby the molding machine. The support surface of this CNC machine consists of an acid etched ($2\,\%$ HNO$_3$ optima grade, $98\,\%$ DI water, $24$\,h) Polyethylene Terephthalate (PET) plate with a vacuum pump being used to hold the disc down, like this avoiding clamps. The inside of the CNC machine's chamber was illuminated with UV light to reduce the static charge of the flakes. Some of the discs were then milled down to a uniform diameter and a hole was cut in the middle of the plate where the sprue\footnote{The sprue refers to the hardened PEN portion that still protruded into the nozzle.} was located. 
This process was performed in order to match the geometry of the chamber of the germanium screening setups \cite{Laubenstein:2017yjj, Brudanin:2017gsu}. After machining, the PEN discs were no longer touched with anything else than a pair of acid etched  pliers with small contact surfaces inside the center hole of the discs.

The individual PEN discs still in the clean room tent, were each sealed in class\,$100$ plastic bags for transportation to the \textit{Modane underground laboratory} (LSM, France) and to the \textit{Laboratori Nazionali del Gran Sasso} (LNGS, Italy), where screening took place. For documentation, the individual bags were numbered and the time of sealing recorded. Transportation happened in large plastic boxes, which were sealed with a radon-impermeable film.

\subsection{Design and processing of the \textsc{Legend}-200 holders}
\label{sec:processing}

\begin{figure}[h]
    \centering
    \includegraphics[width=0.66\textwidth]{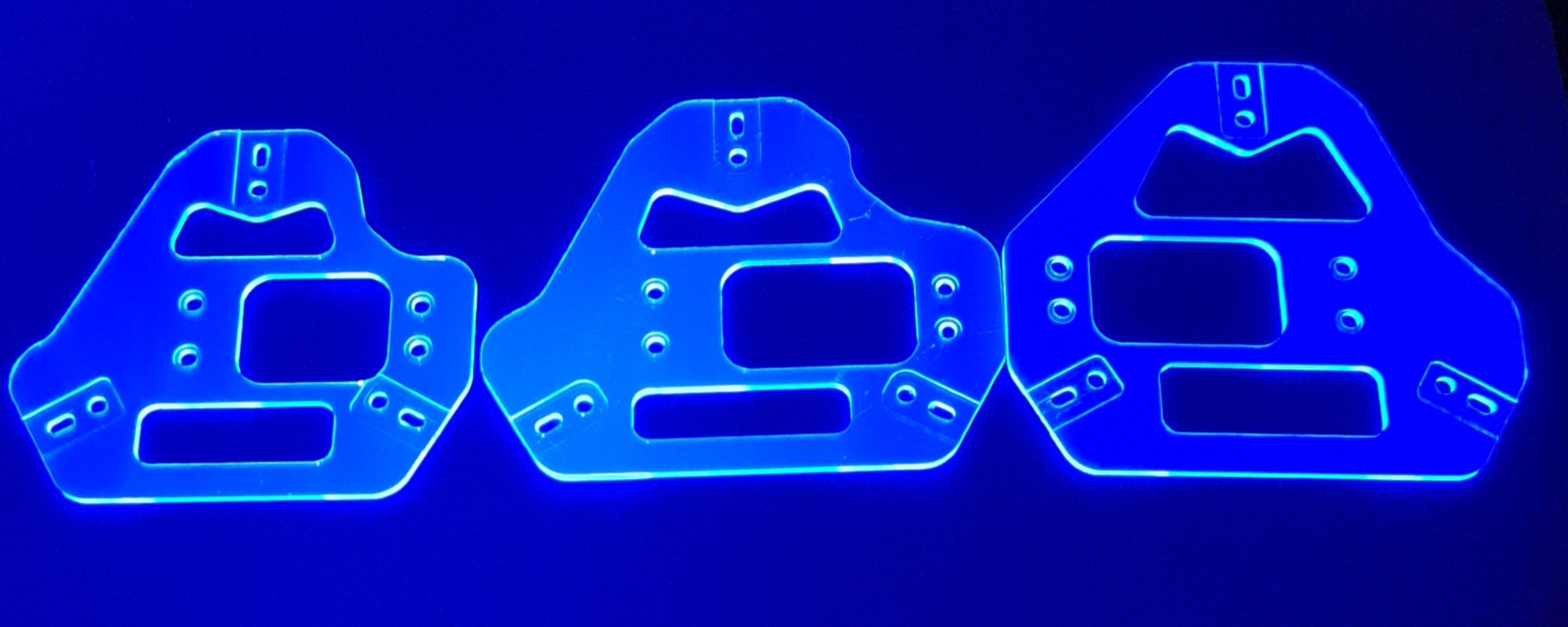}
     \includegraphics[width=0.29\textwidth]{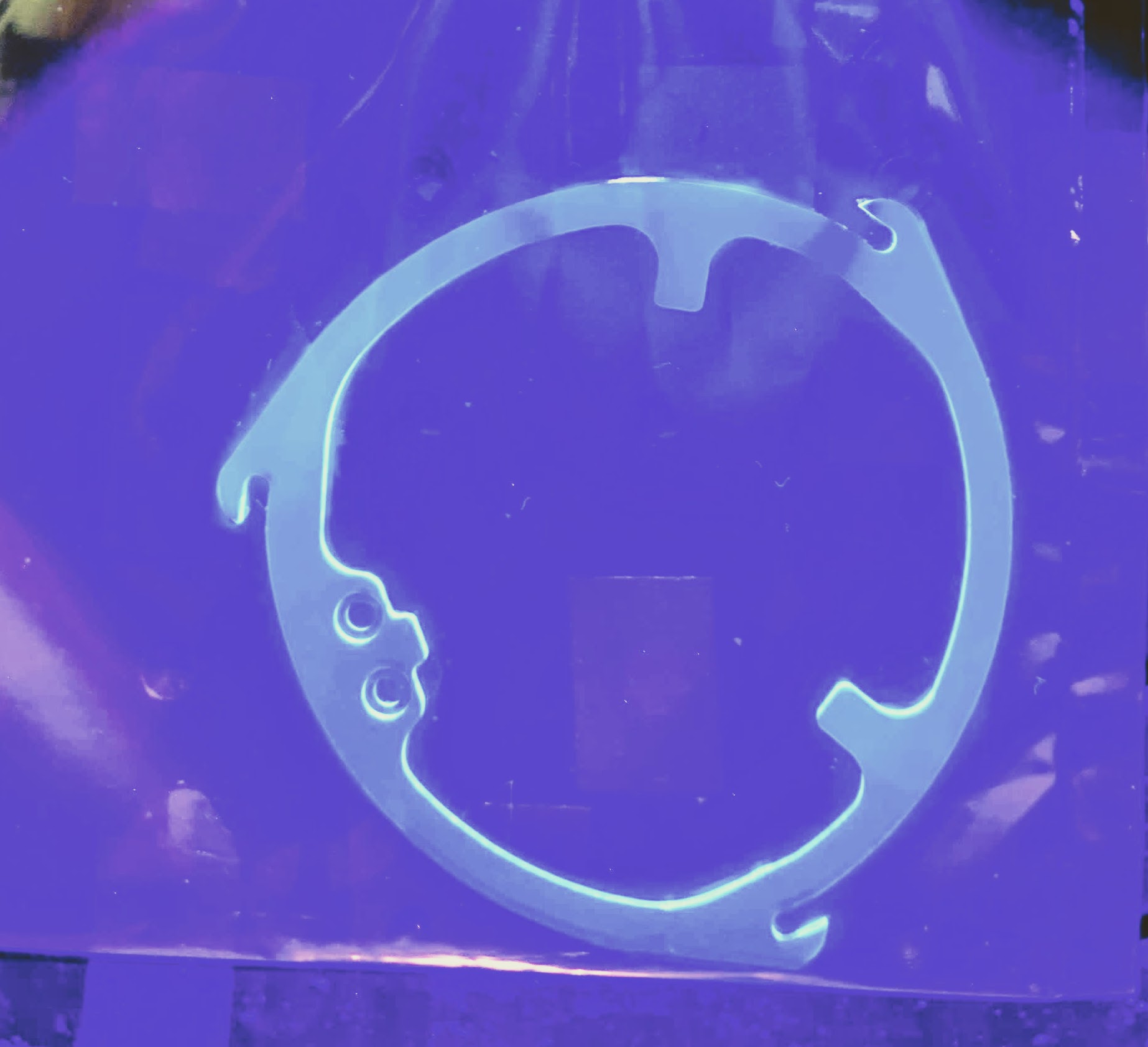}
    \caption{Left: PEN holders geometries required to mount small (Left), medium (Center) and large (Right) Ge detectors for \textsc{Legend}-$200$. Right: Special PEN holder required to place the high voltage on top of some Ge detectors}
    \label{fig:pen_holders} 
\end{figure}

Within \textsc{Legend}-200 a variety of detector geometries will be used, which requires different types of support structures. In particular, it was found that by using three different sizes of PEN holders, all Ge detectors used in \textsc{Legend}-$200$ can be accommodated. In the following, these holder geometries shown in Figure~\ref{fig:pen_holders}~(Left) will be referred to as small, medium and large size. 
In addition, a special PEN holder was designed to mount some detectors that required two PEN holders, one at the bottom and one at the top. The design of this special top holder used to place the high voltage is shown in Figure~\ref{fig:pen_holders}~(Right).
In \textsc{Legend}-200, Ge detectors are mounted in so-called strings where they hang one above the other. Each detector is supported by a holding structure as can be seen in Figure \ref{fig:pgt_detector}. These support structures are composed of a base holder made of PEN (previously Si), and other elements made of copper, ULTEM and PTFE \cite{LEGEND:2021bnm}.

\begin{figure}[h]
    \centering
    \includegraphics[width=0.29\textwidth]{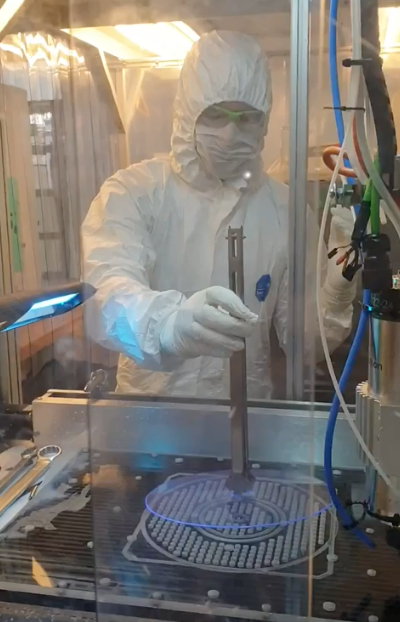}
    \includegraphics[width=0.65\textwidth]{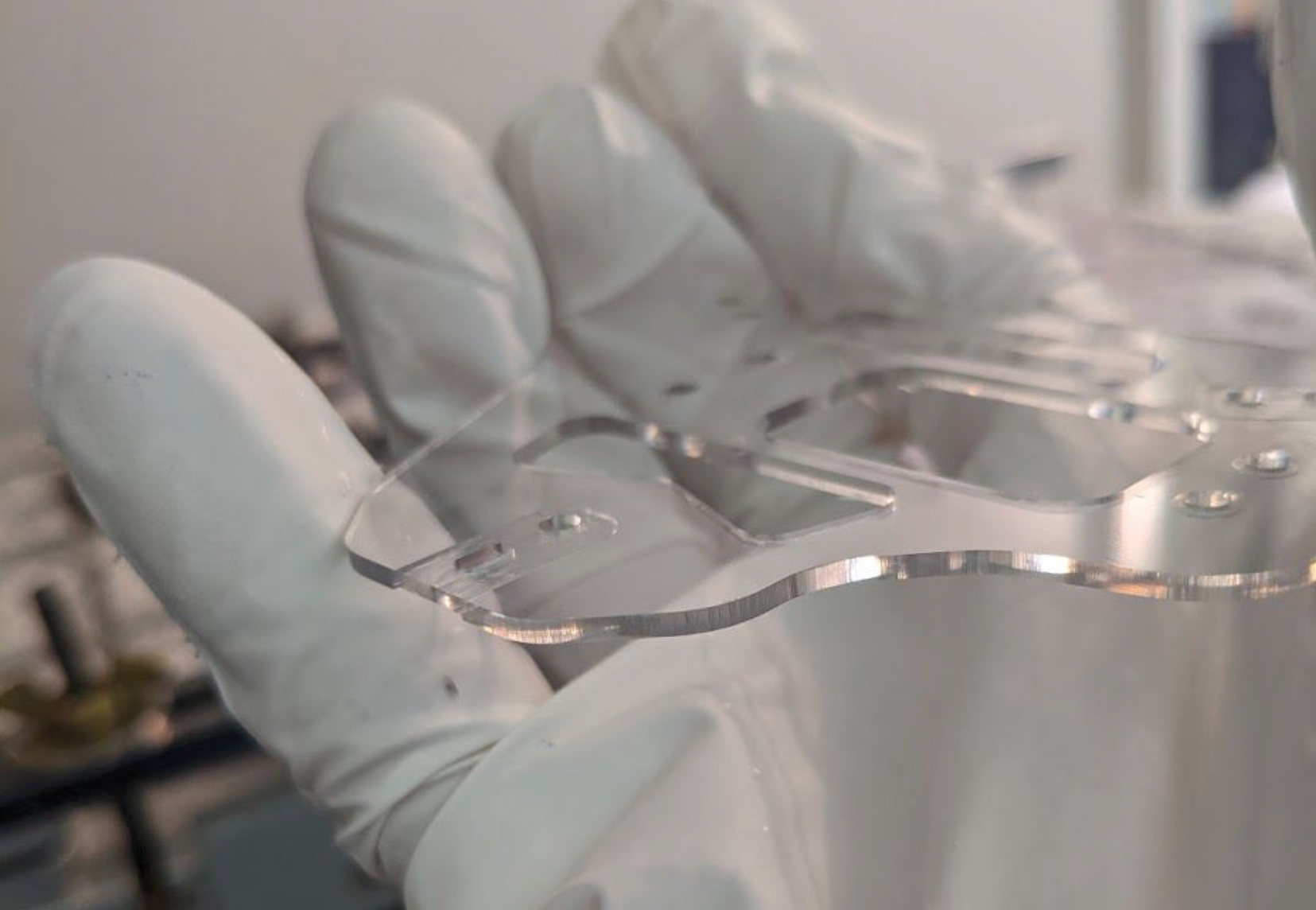}
    \caption{Left: Picture taken while removing a molded PEN disc from the CNC milling machine inside the clean room. Right: Finished PEN holder for the \textsc{Legend}-$200$ experiment machined from a bigger disc (Left). The edges and surfaces are without any additional treatment.}
    \label{fig:holder_in_hand} 
\end{figure}

In order to minimize the mass of the PEN holders and thus to minimize background, mechanical simulations were performed. To this end, a tension load of up to $20$\,N in each string of detectors with a lateral acceleration of the assembly up to $5$\,cm/s$^2$ were assumed. These simulations coupled with a topological optimization using the software \textit{Abaqus $2017$} showed the areas in the designed holders that have minimum stress values and thus could be removed. The results allowed an average mass reduction of $28$\%. Thus, the estimated mass of the holders is $5.3$\,g, $6.3$\,g, and $7.1$\,g corresponding to the small, medium, and large size, respectively. 

Final machining of the optimized \textsc{Legend}-$200$ holders was completed once the screening results were available. Here, the milling machine was setup in a laminar flow box. The operators wore clean room clothing and all parts of the machine were cleaned using $2$-propanol (VLSI grade). In addition, no other machine was running in the room during the time span of the milling process to avoid polluting the air with dust or flakes. MKD/mono-crystalline diamond milling cutters for high-gloss mirror finish from \textit{Karnasch professional tools} were used for the machining. These diamond tools were cleaned following the same procedure as described in Section~\ref{sec:cleaning}. The hardness of the diamond combined with the cleaning protocol guaranteed a minimum risk of introducing radio-impurities during the production.

During the milling process, the PEN disc was held down on a new cleaned poly-carbonate base plate using double-sided tape. Here, the taped parts were then cut away and tape residuals on the base plate were removed after processing each disc. The flakes and dust produced during machining were blown away using a nitrogen line connected directly to a $150$\,l liquid nitrogen dewar. The same nitrogen line was used for cooling the milling cutter during the milling process. The processing of each disc took about $30$\,minutes with $30000 - 33000$\,rpm for the drill and a feed between $480$ and $780$\,mm/min. One disc could fit up to $4$\,medium/large or $8$\,small size holders for \textsc{Legend}-$200$. A finished holder without any additional treatment can be seen in Figure~\ref{fig:holder_in_hand} (right). The edges are transparent and almost no burr is visible. All holders were individually packed and then sent back for another cleaning cycle as described in Section~\ref{sec:cleaning}. 

\section{Radiopurity of produced PEN discs and finished holders}
\label{sec:screening}
The physics goal of the \textsc{Legend}-200 experiment, is to achieve a half-life sensitivity of $10^{27}$\,yr for the $0\nu\beta\beta$-decay in $^{76}$Ge within 5\,years of data taking. To achieve this goal a background index of less than $10^{-4}$\,cts$/(\text{keV}\cdot \text{kg} \cdot \text{yr})$ is required.
Thus, an improvement of one order of magnitude with respect to GERDA background levels (< $10^{-3}$\,cts$/(\text{keV}\cdot \text{kg} \cdot \text{yr})$) is targeted \cite{Avignone:2019phg}. 
In order to accomplish this background index, a maximum radioimpurity of  $1$\,$\mu$Bq/holder can be tolerated \cite{Edzards:2020qdd,Burlac:2021usm,DAndrea:2021gqg}. To demonstrate that the radiopurity achieved in this PEN production meets the requirement for usage in the \textsc{Legend}-$200$ experiment, independent radiopurity measurements were carried out:  HPGe screening, ICP-MS and radon emanation measurements were performed. All measurements made on the cleaned and processed material confirm that the radiopurity is high enough for \textsc{Legend}-200. Moreover, once WLS and scintillation is taking into account, first estimations show that the achieved background levels could satisfy \textsc{Legend}-1000 requirements \cite{LEGEND:2021bnm}.  \textsc{Legend}-1000 aims to reach a half-life sensitivity beyond $10^{28}$\,yr, which needs a  background index of less than $10^{-5}$\,cts$/(\text{keV}\cdot \text{kg} \cdot \text{yr})$.

\subsection{HPGe screening}
\label{sec:hpgescreening}
Different samples of the molded plates were measured using high-purity germanium (HPGe) detectors. One measurement was performed for about $68$\,days at the LNGS with the GeMPI$4$ setup \cite{Laubenstein:2017yjj}, using $14.315$\,kg of PEN.
Another measurement was done with the OBELIX detector at the LSM \cite{Brudanin:2017gsu}, using $5.231$\,kg of PEN. This screening took about $79$\,days.

\begin{table} [h]
\begin{center}
\begin{tabular}{ |c|c c|c c|c c|c c|}
 \hline
 \multirow{4}{*}{}   & \multicolumn{2}{c|}{Raw TN-$8065$S} & \multicolumn{2}{c|}{Discs}  & \multicolumn{2}{c|} {Discs}  & \multicolumn{2}{c|} {L200 holders} \\
                            & \multicolumn{2}{c|}{GeMPI$4$ at LNGS} & \multicolumn{2}{c|}{ GeMPI$4$ at LNGS   }          & \multicolumn{2}{c|} {OBELIX at LSM} & \multicolumn{2}{c|} {GeMPI$3$ at LNGS} \\ \cline{2-9}
                            & \multicolumn{2}{c|}{-} & \multicolumn{2}{c|}{\myvec{$14.315$\,kg \\ $68$\,days}}        & \multicolumn{2}{c|} {\myvec{$5.231$\,kg \\ $79$\,days}}  & \multicolumn{2}{c|} {\myvec{$1.07$\,kg \\ $68$\,days}} \\ \cline{2-9}
                            \hline \hline
             
 $^{228}$Ra  & \multicolumn{2}{c|}{$< 0.15$\,mBq/kg} & \multicolumn{2}{c|}{ ($92 \pm 25$)\,$\mu$Bq/kg } & \multicolumn{2}{c|}{ ($107 \pm 38$)\,$\mu$Bq/kg}  & \multicolumn{2}{c|}{ $<0.46$\,mBq/kg} \\ \hline 
 $^{228}$Th  & \multicolumn{2}{c|}{($0.23 \pm 0.05$)\,mBq/kg} & \multicolumn{2}{c|}{ ($32 \pm 16$)\,$\mu$Bq/kg } & \multicolumn{2}{c|}{ ($67 \pm 18$)\,$\mu$Bq/kg } & \multicolumn{2}{c|}{  $<0.48$\,mBq/kg} \\ \hline
 $^{226}$Ra  & \multicolumn{2}{c|}{($0.25 \pm 0.05$)\,mBq/kg} & \multicolumn{2}{c|}{ ($60 \pm 15$)\,$\mu$Bq/kg } & \multicolumn{2}{c|}{ ($76 \pm 22$)\,$\mu$Bq/kg } & \multicolumn{2}{c|}{  $<0.36$\,mBq/kg} \\ \hline \hline
 $^{234}$Th  & \multicolumn{2}{c|}{$< 11$\,mBq/kg} & \multicolumn{2}{c|}{$< 1.9$\,mBq/kg }                 &\multicolumn{2}{c|}{ -}  & \multicolumn{2}{c|}{  $<5.8$\,mBq/kg} \\ \hline
$^{234}$Pa   & \multicolumn{2}{c|}{$< 3.4$\,mBq/kg} & \multicolumn{2}{c|}{$< 1.7$\,mBq/kg}                  & \multicolumn{2}{c|}{ -}  & \multicolumn{2}{c|}{  $<7.0$\,mBq/kg} \\ \hline
 $^{235}$U   & \multicolumn{2}{c|}{$< 0.066$\,mBq/kg} & \multicolumn{2}{c|}{$ < 56$\,$\mu$Bq/kg }             & \multicolumn{2}{c|}{ -}  & \multicolumn{2}{c|}{  $<0.22$\,mBq/kg} \\ \hline
 $^{40}$K    & \multicolumn{2}{c|}{($1600 \pm 400$)\,mBq/kg} & \multicolumn{2}{c|}{$ < 0.24$\,mBq/kg}                & \multicolumn{2}{c|}{$ <0.567 $\,mBq/kg}   & \multicolumn{2}{c|}{  $<4.1$\,mBq/kg} \\ \hline 
  $^{137}$Cs & \multicolumn{2}{c|}{$< 0.057$\,mBq/kg} & \multicolumn{2}{c|}{$ < 0.15$\,$\mu$Bq/kg}            & \multicolumn{2}{c|}{ -}  & \multicolumn{2}{c|}{  $<0.091$\,mBq/kg} \\ \hline
\end{tabular}
\end{center}
\caption{Radiopurity results as achieved for molded PEN components using commercially available PEN granulate, subsequently cleaned. Two sets of molded PEN samples were measured at LNGS and LSM underground laboratories. Furthermore, measurements of uncleaned granulate \cite{Efremenko:2019xbs} and the finished holders for \textsc{Legend}-$200$ (L$200$) are listed. Not all listed radionuclides were analyzed in the OBELIX setup. Limits are given with $90\%$ confidence level.}
\label{tab:PEN:rp}
\end{table}

The results of all HPGe screening measurements are reported in Table\,\ref{tab:PEN:rp} for raw PEN granulate, for two independent measurements of the injection molded discs and for the final PEN holders. 
Clearly, the radio-impurities measured with the granulate delivered by DuPont was significantly reduced thanks to the cleaning procedure. Notably, it could be confirmed that the high $^{40}$K signal was due to surface contamination. 
More important, the radiopurity achieved during this first production run satisfy \textsc{Legend}-$200$ requirements. 

It should be noted here that the mass of the measured final holders ($1.07$\,kg) is much lower compared to the original big discs which were designed to fill the chamber. Hence, no comparable exposure could be achieved and only limits were obtained in this final measurement. These results confirm nevertheless that no noticeable contamination has been introduced during the post-production.   

\subsection{ICP-MS}
\label{sec:icp-ms}
Two finished holder samples were also sent for inductively coupled plasma mass spectrometry (ICP-MS) at the \textit{Pacific Northwest National Laboratory} (PNNL). The ICP-MS method allows to reach a $\mu$Bq sensitivity on $^{232}$Th and $^{238}$U. The two holders were subsampled in triplicate using new and cleaned stainless-steel scissors. The subsamples went through a cleaning procedure including two subsequent leaching in $5$\% optima grade HNO$_3$ for two hours. The leachates were transferred into validated PFA Savillex vials and directly analyzed on an Agilent $8900$ ICP-MS, to ensure any surface contamination from transportation and subsampling had been properly removed. An external calibration standard was used for quantitation of $^{232}$Th and $^{238}$U. After the contamination in the leachates was measured to be low enough, within instrumental background, the subsamples were placed into validated microwave vessels for digestion.  Table~\ref{tab:icpms-leaches} reports the measured values for each leach. It was found that in general the level of $^{232}$Th and $^{238}$U decreased from the first leach to the second leach. These levels could be representative of the remaining contamination from the parts fabrication, handling and/or from the subsampling.
It should be noted here, that for the ICP-MS measurement of the leachates the actual sample mass in the leach is not known. For the results, the total mass of the sample was used. The values given can therefore be used to understand the trend but do not accurately quantify the amount of $^{232}$Th and $^{238}$U.

\begin{table} [h]
\begin{center}
\begin{tabular}{ |c|c|c c|c c|}
 \hline
 \textbf{Sample} & & \multicolumn{2}{c|}{\textbf{$^{232}$Th}} & \multicolumn{2}{c|}{\textbf{$^{238}$U}} \\
 \textbf{Weight [g]} &  & \textbf{ppt} & \textbf{$\pm$ inst} &\textbf{ ppt} & \textbf{$\pm$ inst} \\ \hline\hline
 \myvec{PEN \#$1$ \\ $0.7907$} & \myvec{leach \#$1$ \\ leach \#$2$} & \myvec{$0.071$ \\ $0.004$} & \myvec{$0.013$ \\ $0.005$} & \myvec{$0.09$ \\ $0.0068$} & \myvec{$0.04$ \\ $0.0017$} \\ \hline
 \myvec{PEN \#$2$ \\ $0.7136$} & \myvec{leach \#$1$ \\ leach \#$2$} & \myvec{$0.10$ \\ $0.0045$} & \myvec{$0.03$ \\ $0.0015$} & \myvec{$0.031$ \\ $0.007$} & \myvec{$0.011$ \\ $0.002$} \\ \hline 

\end{tabular}
\end{center}
\caption{Results of the ICP-MS measurement of the leachates for $^{232}$Th and $^{238}$U, normalized to the mass of the sample. Absolute detection limits were calculated as $3$ times the standard deviation of the process blanks and the instrumental precision for a given measurement is given as $\pm$ inst.}
\label{tab:icpms-leaches}
\end{table}

A known amount of non-natural $^{229}$Th and $^{233}$U was then spiked into each replicate and process blank. Subsamples were digested at $250^\circ$C in concentrated optima grade HNO$_3$ and then transferred to validated Savillex vials. The subsamples were dried down on a hotplate, reconstituted in $2$\% optima grade HNO$_3$, and analyzed on an Agilent $8900$ ICP-MS. Quantitation of $^{232}$Th and $^{238}$U was performed using isotope dilution mass spectrometric methods.
Table~\ref{tab:icpms-final} reports the limits obtained from the PEN subsamples. Reported results are given in ppt for $^{232}$Th and $^{238}$U. Absolute detection limits were calculated as $3$ times the standard deviation of the process blanks. All measurements for $^{232}$Th and $^{238}$U were below relative detection limits, which were calculated by dividing the absolute detection limits by the subsample masses. These results combined with the gamma spectroscopy measurements confirm that an excellent radiopurity has been achieved. In addition, these ICP-MS results suggest that remaining surface contamination can be removed and that values reported in Table~\ref{tab:PEN:rp} should be considered as upper limits for the bulk contamination.

\begin{table} [h]
\begin{center}
\begin{tabular}{ |c|c|c|c|c|}
 \hline
  & & & \textbf{$^{232}$Th} & \textbf{$^{238}$U} \\
 \textbf{Description} & \textbf{Replicate} & \textbf{Sample wt [g]} & \textbf{ppt} &\textbf{ ppt} \\ \hline\hline
 PEN \#$1$ & \myvec{$1$ \\ $2$ \\ $3$} & \myvec{$0.2745$ \\ $0.2481$ \\ $0.2655$} & \myvec{$<1.13$ \\ $<1.30$ \\ $<1.22$} & \myvec{$<1.98$ \\ $<2.08$ \\ $<1.94$} \\ \hline 
 PEN \#$2$ & \myvec{$1$ \\ $2$ \\ $3$} & \myvec{$0.2784$ \\ $0.1408$ \\ $0.2940$} & \myvec{$<1.02$ \\ $<2.02$ \\ $<0.97$} & \myvec{$<1.81$ \\ $<3.57$ \\ $<1.71$} \\ \hline 
\end{tabular}
\end{center}
\caption{Results of the ICP-MS measurement of the the PEN subsamples for $^{232}$Th and $^{238}$U. Absolute detection limits were calculated as 3 times the standard deviation of the process blanks.}
\label{tab:icpms-final}
\end{table}

\subsection{Radon emanation}
\label{sec:radom-emanation}

Complementary radiopurity information can be obtained by means of high sensitivity Radon emanation measurement techniques \cite{Wojcik:2007zza,Wojcik:2017}. 
Five finished PEN holders were screened using this method at the Institute of Physics of the \textit{Jagiellonian University} in Krakow.  Given the small size of the samples, only limits of less than 12~$\mu$Bq/holder  and less than 16~$\mu$Bq/holder for $^{222}$Rn and $^{220}$Rn respectively were derived. These limits were obtained from two measurements carried out using the five holders placed into a 50 liter chamber.

\section{Optical properties}
\label{sec:optical_properties}

Besides the excellent mechanical properties of PEN at cryogenic temperatures \cite{Efremenko:2019xbs} and the very good radiopurity achieved during this first production for \textsc{Legend}-200, the wavelength shifting capabilities and scintillation properties provide major advantages. Thus, the optical properties are very important and, among others, include the surface quality, light yield and attenuation length. Custom-made scintillators as described in this paper, could have different properties depending on various settings in the production. In order to classify and compare production batches, two test stations were designed. In addition, the surface of the molded PEN discs has been characterized using different methods. These results showed that high quality surfaces have been obtained. 
\begin{table}[h]
    \centering
    \begin{tabular}{ |c|c| } 
    \hline
 Light output\footnote{The light output was obtained exciting PEN and reference samples using electrons from a $^{207}$Bi source. The light output was then compared to the well known reference samples of type EJ-200 from Eljen Technology  of the same dimensions. 
 The quoted light yield of EJ-200 is 10000$\pm$200 photons per MeV. Thus 
 two stacked square samples of 30$\times$30$\times$1.7~mm$^{3}$ were used for both scintillators. 
 Two samples were necessary, in order to completely absorb the high energy electrons from the $^{207}$Bi source ($\sim$ 1 MeV).
 These samples were coupled to five 1 inch photomultiplier tubes (PMTs) using optical grease (EJ-550). Only one 30$\times$30~mm$^{2}$ side was free to place the radioactive source. } & 30$\pm$2\% of EJ-200 \\ 
 \hline
 Light yield\footnote{The light yield is estimated using Geant4 simulations since a direct comparison between PEN and EJ-200 provides only a lower limit of 3000 photons per MeV. These simulations are needed to take into account attenuation effects. The bulk absorption of the EJ-200 reference samples is of the order of 4 meters, while for PEN it is only a few centimeters. Therefore, attenuation effects are expected to be important even for small PEN samples while for EJ-200 are negligible. } & 5500$\pm$550 photons/MeVe$^{-}$  \\
 \hline
 Time constant & 25.3 $\pm$0.2 ns \\ 
 \hline
 Bulk absorption & $\sim$60 mm at 450 nm  \\ 
 \hline
 Emission spectrum maximum & 440 $\pm$ 5 nm  \\ 
 \hline
\end{tabular}
    \caption{Main optical and scintillation properties of \textsc{Legend}-200 PEN material. \cite{pen_optics,Abt:2020pwk}}
    \label{tab:optics_pen}
\end{table}

\noindent A detailed description of the setups and the results of the optical properties of this \textsc{Legend}-200 PEN production, as well as estimation of self-vetoing capabilities of the PEN holders is under preparation in reference \cite{pen_optics}.  The main optical properties are summarized in table \ref{tab:optics_pen}.
\section{PEN as detector holder for HPGe detectors}
\label{sec:tests}
In order to qualify the PEN detector holders for usage in \textsc{Legend}-200, two proof of concept tests have been performed. In both tests, Ge detectors were mounted on PEN holding plates from the production run described in Section~\ref{sec:production} and operated in LAr. The goal of both tests was to qualify the mounting process and demonstrate the stability of the holders at low temperatures under realistic loads and close to final operational conditions. Also the baseline of the HPGe detectors mounted on PEN detector holders was monitored in order to verify the required stability of the leakage current over longer time spans. An increase of the baseline can be traced back to an induced leakage current via the resistor in the front-end electronics. \begin{figure}[h]
    \centering
    \includegraphics[width=0.5\textwidth]{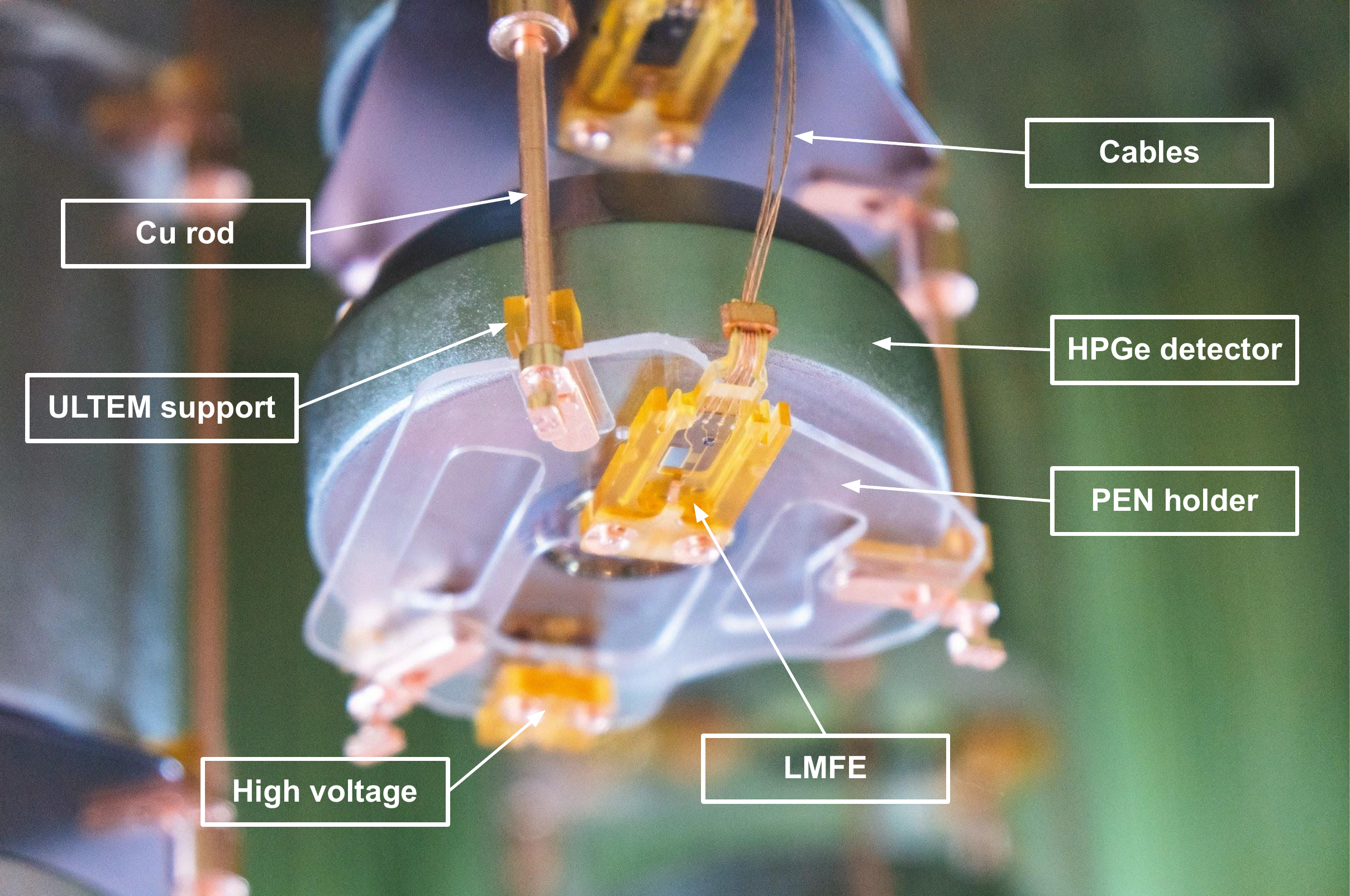}
    \includegraphics[width=0.45\textwidth]{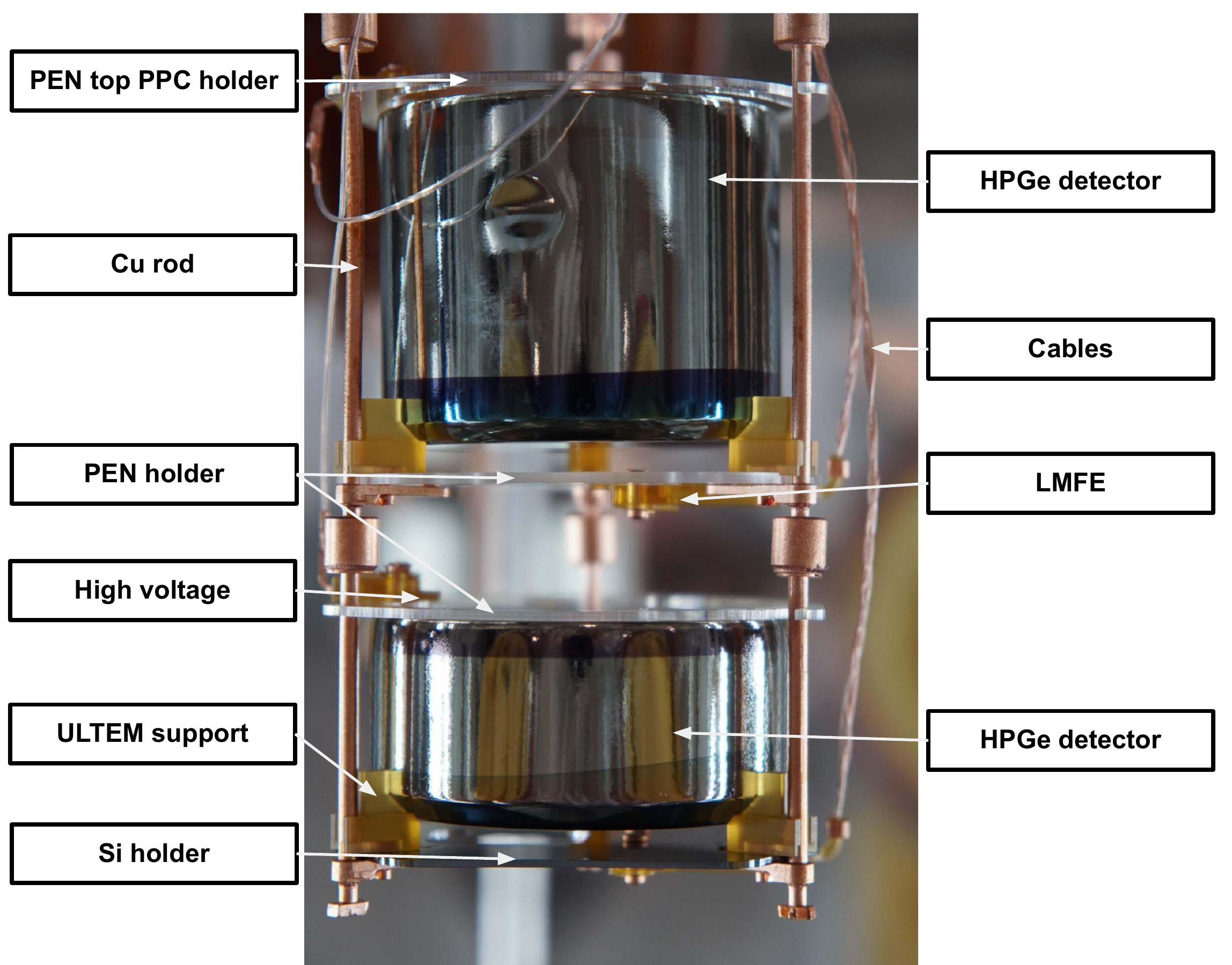}
    \caption{Left: Bottom view of Ge detector string in the \textsc{Gerda} setup during the \textit{Post-\textsc{Gerda} test}. Detector in focus is mounted on PEN holder with front-end electronics attached. Right: Lateral view of a string deployment test at LNGS using PPC type detectors, requiring a top PEN holder to place the high voltage connection.}
    \label{fig:pgt_detector}
\end{figure}

The first test was performed in the underground SCARF setup at TU Munich, Germany \cite{Csathy:2016wdy}. Two small detectors and the front end electronics were mounted on PEN holders. As PEN is more flexible than previously used materials, the parameters for the wire bonding process had to be adapted. For both detectors the bonding worked as usual and no problems due to the flexibility or other mechanical properties of PEN arose. Then the detectors were operated in SCARF in LAr for 10 days. The measured fluctuations in the leakage current were of the order of pA. The change within the measured time span is according to expectations verifying that there is no effect of the PEN holders on the leakage current of the Ge detectors.

\begin{figure}[h]
    \centering
    \includegraphics[width=\textwidth]{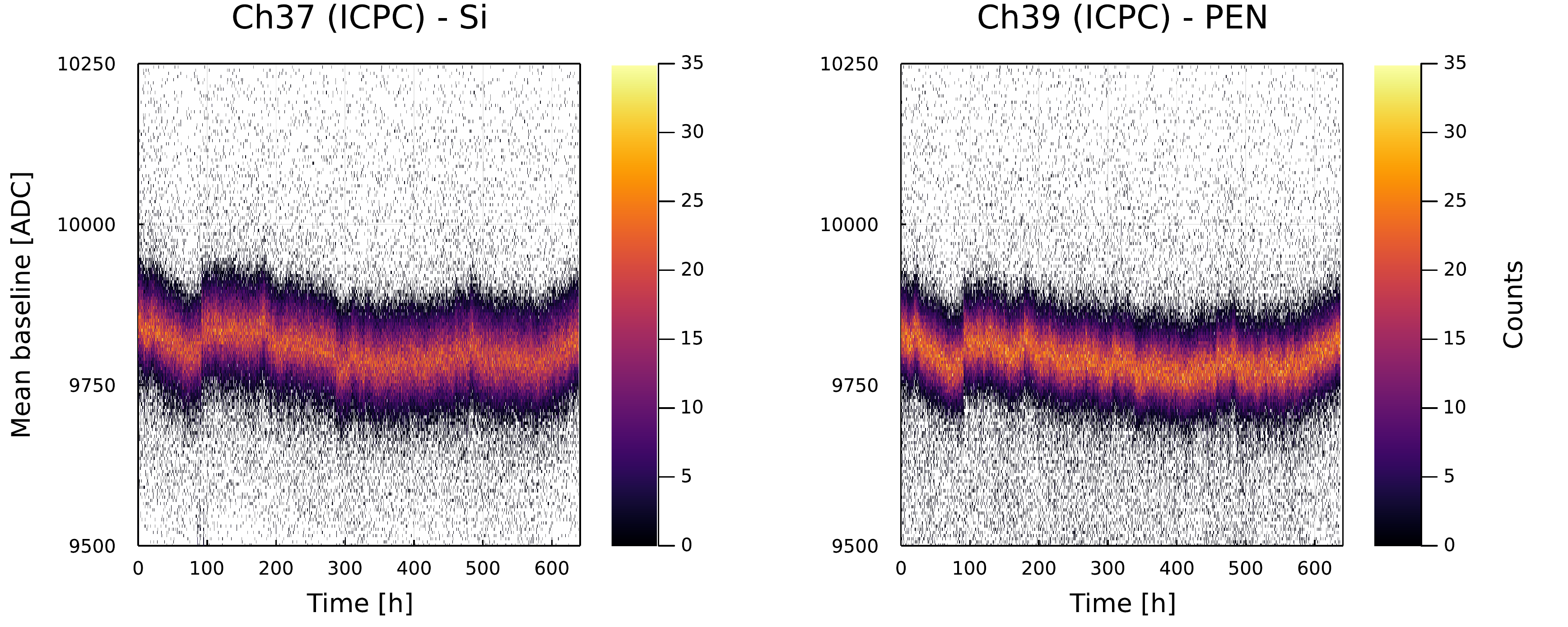}
    \caption{Mean baseline over time of two ICPC Ge detectors operated during the PGT. One detector was mounted on Si (left) and one on PEN (right).}
    \label{fig:pgt_baseline}
\end{figure}

The second test was started in February $2020$ during the \textit{Post-\textsc{Gerda} test} (PGT). During the PGT several detector strings were deployed using the \textsc{Gerda} infrastructure in order to validate \textsc{Legend}-200 electronics and other components. Around 40\% of Ge detectors were mounted using PEN holders. This allowed a direct comparison of detectors mounted on PEN holders to detectors mounted with conventional Si holders as used previously in \textsc{Gerda}. A picture of the lower part of a detector string during the PGT can be seen in Figure~\ref{fig:pgt_detector}~(Left). The detector in focus is mounted on a PEN holder. Figure~\ref{fig:pgt_detector}~(Right) shows the lateral view of a string test deployment at LNGS using PPC detectors. 
The longest run lasted about $27$\,days. Figure \ref{fig:pgt_baseline} shows the mean value of the baseline for two detectors for that time period. Here, two new inverted-coaxial point-contact (ICPC) Ge detectors are compared, one of which was mounted on Si (Ch$37$, left) and one on PEN (Ch$39$, right). For the determination of the mean value of the baseline, the first $1000$\,samples of the waveforms were averaged. There was a comparable behavior and no significant variations were observed in both data sets. There were also no significant differences in the energy resolution.
Thus, for the $2614.5$\,keV $\gamma$-line of $^{208}$Tl the achieved resolutions (full width at half maximum, FWHM) for a detector mounted on a Si or PEN holder are $2.43$\,keV and $2.61$\,keV,  respectively \cite{LEGEND:2021bnm, FrankThesis}. These deviations are in line with expectations for the type of detector used.
Extrapolated to the $Q_{\beta\beta}$ of the $0\nu\beta\beta$-decay in $^{76}$Ge (2039 keV), the detector mounted on a Si holder shows an energy resolution of $2.17$\,keV and the one mounted on a PEN holder of $2.33$\,keV. 

\section{Conclusions \& Outlook} \label{sec:conclusions}

The first production series of low-background components made of PEN for \textsc{Legend}-200 was successfully finished. The introduction of an intensive cleaning cycle of the raw granulate removed most of the surface contaminants found in previous measurements. Furthermore, it could be shown that remaining impurities on the surface can be removed by leaching. The current components thus have an activity of less than $1$\,$\mu$Bq per holder, meeting the \textsc{Legend}-200 requirements. In addition, 
R\&D is ongoing exploring self-synthesis of the PEN raw material, which  could reduce even more the achieved radiopurity.

Preliminary estimations suggest a light yield of about $5500$\,photons per MeV and a bulk absorption of around $60$\,mm at $500$\,nm. In addition,
the application of PEN as a low background component was also tested in two setups under close to final operational conditions. In both cases, Ge detectors mounted with PEN holders were operated in LAr for a period of time on the order of months. No complications arose either in the installation of the electronics or in operation.
Leakage current in the Ge detectors was  not affected by PEN and it was found to be comparable with previous measurements. 
Direct comparisons with Ge detectors during the post-\textsc{Gerda} tests showed no differences between detectors mounted with Si or PEN holders. This period of this time is a reasonable representation of the stability needed for a 10+ year experiment.

\textsc{Legend}-$200$ will be commissioned in the further course of $2021$. This will be the first time that custom-made low-background PEN scintillating components are used in a rare event search experiment.

\acknowledgments
This work was supported by the Ministry of Industry and Trade of the Czech Republic under the Contract Number FV30231. Research sponsored by the Laboratory Directed Research and Development Program of Oak Ridge National Laboratory, managed by UT-Battelle, LLC, for the U.S. Department of Energy. This material is based upon work supported by the U.S. Department of Energy, Office of Science, Office of Nuclear Physics, under Award Number DE-AC05-00OR22725. Felix Fischer and Luis Manzanillas are supported by the Deutsche Forschungsgemeinschaft (SFB1258). Connor Hayward is supported by the Excellence Cluster Universe. The authors would like to  thank M. Laubenstein (LNGS) G. Zuzel (Cracow) and E. Hoppe (PNNL) for
their invaluable help by delivering the screening results of the PEN
samples.

\bibliographystyle{unsrt}
\bibliography{bib}

\end{document}